\input harvmac
\Title{IFP-805-UNC}{Zeroes of the Neutrino Mass Matrix}
\centerline{Paul H.  Frampton,${}^{(a)}$ \  Sheldon L. Glashow$\,{}^{(b)}$
and Danny Marfatia$\,{}^{(b)}$}
\bigskip

\centerline{${}^{(a)}$\ \it Department of Physics and Astronomy}
\centerline{$\phantom{{}^{(a)}}$ \ \it University of North Carolina, Chapel
Hill, NC 27599-3255}
\bigskip

\centerline{${}^{(b)}$\ \it Department of Physics}
\centerline{$\phantom{{}^{(a)}}$ \ \it Boston University, Boston, MA 02215}

\vskip .5in

We assume there to be precisely three left-handed neutrino states whose
Majorana masses are generated by an unspecified mechanism.  Were CP conserved,
the symmetric neutrino mass matrix $\cal M$ would be real and all six of its
distinct entries could be experimentally determined.
 But CP is not conserved so that $\cal M$
is likely to be complex. As a result, not all nine of its
convention-independent real parameters can be determined without an appeal to
theory. Thus we examine the possibility that a restricted class of neutrino
mass matrices may suffice to describe current data, namely those complex
symmetric matrices several of whose entries vanish.  We find that there are
seven  acceptable textures with two independent zeroes, and we explore their
contrasting phenomenological implications. Textures with more than two
independent zeroes appear to be excluded by experiment.

\Date{01/02}

\newsec{Introduction} Quark masses and mixings are described by two $3\times
3$ matrices which involve a total of ten convention-independent real
quantities: four Kobayashi-Maskawa parameters and six positive quark
masses. The lepton sector requires only three additional parameters (the
charged lepton masses) in the minimal---and manifestly incomplete---standard
model wherein neutrinos are massless.

In this paper we assume that neutrino phenomenology can be formulated in
terms of three left-handed neutrino states with a complex symmetric Majorana
mass matrix $\cal M$. We do not commit ourselves to any particular mechanism
by which neutrino masses are generated. Of the twelve real parameters
characterizing $\cal M$, three are arbitrary phases of the three flavor
eigenstates. Thus the neutrino sector involves nine convention-independent
parameters. The two squared-mass differences and the four Kobayashi-Maskawa
analogs are likely to be measureable in practice. So also is the quantity
$\vert {\cal M}_{ee}\vert$ through the search for neutrinoless double beta
decay. There remain two parameters that are measureable in principle but
apparently not in practice. We arrive at the dreadful conclusion that 
{\it no presently conceivable set of feasible experiments can 
fully determine the neutrino mass matrix.}

 This is the
context and the justification for our discussion of
a conjectured simplicity  in the structure
of the neutrino mass matrix.   In particular
we propose  that $\cal M$, expressed in the $(e, \,\mu,\, \tau)$ flavor
basis, may have 
 several vanishing entries.  The number $n$ of independent zeroes
takes into account the symmetry of the mass matrix: it counts instances that
${\cal M}_{ij}=0$ where $i\ge j$. We find  that currently available
 data disfavor all cases with  $n\ge 3$, but that there are seven
empirically tolerable textures with $n=2$.

We begin by recalling the standard notation~\ref\rgg{{\it E.g.,} H.
  Georgi and S.L. Glashow, Phys. Rev. {\bf D61,} 097301 (2000).}\
 for three-flavor neutrino
oscillations. With an appropriate choice of the phases of the three flavor
  eigenstates, we may express $\cal M$ as follows:
\eqn\eM{{\cal M}= U^*\,D\,U^\dagger}
where $U$ is the neutrino analog to the Kobayashi-Maskawa matrix
expressing flavor eigenstates in terms of mass eigenstates:
\eqn\eKM{\pmatrix{\nu_e\cr\nu_\mu\cr\nu_\tau\cr}=
\pmatrix{c_2c_3&c_2s_3&s_2\,e^{-i\delta}\cr
-c_1s_3-s_1s_2c_3\,e^{i\delta}&+c_1c_3-s_1s_2s_3\,e^{i\delta}&s_1c_2\cr
+s_1s_3-c_1s_2c_3\,e^{i\delta}&-s_1c_3-c_1s_2s_3\,e^{i\delta}&c_1c_2\cr}\,
\pmatrix{\nu_1\cr\nu_2\cr\nu_3\cr}}
with $s_i$ and $c_i$ standing for sines and cosines of $\theta_i$.
The remaining five parameters appear in the diagonal matrix
\eqn\eD{D=\pmatrix{m_1&0&0\cr0&m_2&0\cr 0&0&m_3}}
where $m_1$ and $m_2$,  in general, are  complex numbers while $m_3$ can be
taken to be real and non-negative with no loss of generality.
In the  analysis to follow
 we  make the following hypotheses, all of which are  favored 
(if not yet established) by
current experimental data:
\medskip

(1) For solar neutrinos, we assume
 oscillations to be  large, but we also assume that they are measurably 
non-maximal~\ref\rgg{{\it E.g.,} J.N. Bahcall,
 M.C. Gonzalez-Garcia and C. Pena-Garay,
{\tt hep-ph/0111150}\semi
and experimental references cited therein.}.
 In particular, we take
$0.6\le \sin^2{2\theta_3}\le 0.96$,
with 0.8 as a best-fit value. The relevant squared-mass difference 
$\big\vert \vert m_1\vert^2-\vert m_2\vert^2\big\vert$ is
taken to be  $\Delta_s\approx  5\times 10^{-5}~\rm eV^2$.
\medskip

(2) For atmospheric neutrinos, we  assume  oscillations to be  large 
(possibly maximal) and dominantly of the form $\nu_\mu\rightarrow \nu_\tau$.
In particular, we take $\tan^2{\theta_1}\simeq 1$~\ref\ratm{Y. Fukuda
{\it et al.} [Super-Kamiokande Collaboration], Phys. Rev. Lett. {\bf 81},
1562 (1998).}.
The relevant squared-mass difference $\big\vert\vert m_{1,2}\vert^2-
\vert m_3\vert^2\vert\big\vert$ is
taken to be
$\Delta_a\approx  3\times 10^{-3}~\rm eV^2$.
\medskip
(3) For the subdominant angle $\theta_2$ which controls atmospheric
    $\nu_\mu\leftrightarrow \nu_e$ oscillations
we assume $\sin^2{2\theta_2}\le 0.1$ in accordance with CHOOZ
data~\ref\rCHOOZ{M Apollonio {\it et al.,} Phys. Lett. {\bf B466,} 
415 (1999).}. 

\medskip

\noindent
It will be useful to define the ratio of squared-mass differences, whose
estimated value is:
\eqn\eR{ R_\nu \equiv {\Delta_s\over \Delta_a}\approx  2\times 10^{-2}.}

\newsec{Seven  Two-Zero Textures}

With the above hypotheses and notation, we turn to the question of which
two independent entries of $\cal M$ can vanish in the basis wherein the
charged lepton mass matrix is diagonal.
 Of the fifteen logical
possibilities, we find   just seven to be 
 in accord with our empirical hypotheses.  We discuss them
individually, with the 
non-vanishing
entries  in each case denoted by $X$'s. Our results are presented to leading
order in the small parameter $s_2$.
 We begin with a texture in which ${\cal M}_{ee}={\cal M}_{e\mu}=0$:

\vfill\eject

\noindent{Case $A_1$: ~~
$ \pmatrix{0&0&X\cr 0&X&X\cr X&X&X\cr}$}
\medskip

\noindent
For this texture,
 ${\cal M}_{ee}=0$ so that  the amplitude for no-neutrino double beta
 decay vanishes to lowest order in neutrino masses. If Case $A_1$
 is realized in nature,
 the neutrinoless process simply cannot be detected. There is even more to
 say.
Using Eqs.~\eM--\eR\  we find:
\eqna\eA
$$\eqalignno{
m_1\simeq +(s_2t_1t_3)\,m_3\,e^{i\delta}\,,~~~&~~~
m_2 \simeq -(s_2t_1/t_3)\,m_3\,e^{i\delta}\,,&\eA a\cr
R_\nu\simeq  s_2^2\,t_1^2&\, \big\vert t_3^2-1/t_3^2\big\vert\,,
&\eA b \cr}$$
where $t_i$ stands for $\tan{\theta_i}$. Two of the squared neutrino masses
are suppressed relative to the third by a factor of $s_2^2$.
As a result
 we find that 
 $s_2$ can lie  close  to
  its present experimental upper limit. This prediction will become more
precise when $\theta_3$ is better measured. 
However,
 the CP-violating parameter $\delta$ is entirely 
unconstrained. For case $A_1$  the subdominant 
angle $\theta_2$ is likely to be
 measureable  and neutrinos may display observable  CP violation.  
\bigskip\medskip

\noindent{Case $A_2$: ~~
$ \pmatrix{0&X&0\cr X&X&X\cr 0&X&X\cr}$}
\medskip
\noindent  This texture, with ${\cal M}_{ee}={\cal M}_{e\tau}=0$,     is
described by Eqs.~\eA{a,b}\ with $t_1$ replaced by $-1/t_1$. Its
phenomenological consequences are nearly the same  as those of Case $A_1$.
\bigskip\medskip

\noindent{Case $B_1$: ~~
$ \pmatrix{X&X&0\cr X&0&X\cr 0&X&X\cr}$}
\medskip
\noindent With  ${\cal M}_{\mu\mu}={\cal M}_{e\tau}=0$, we find
 an acceptable solution if and only if 
$\vert s_2 \,\cos{\delta}\tan{2\theta_1}\vert \ll 1$, in which case we obtain:
\eqna\eB
$$\eqalignno{
m_1&\ \simeq -\big(t_1^2+s_2\,(e^{-i\delta}t_1+e^{i\delta}/t_1)/t_3\big)\,m_3
\,,&\eB a\cr
m_2&\ \simeq -\big(t_1^2-s_2\,(e^{-i\delta}t_1+e^{i\delta}
/t_1)\,t_3\,\big)\,m_3\,,
&\eB b\cr
R_\nu &\ \simeq \vert s_2\,\cos{\delta}\, \tan{2\theta_1}(t_3+1/t_3)\vert\,.
&\eB c\cr
}$$
The three neutrinos are nearly degenerate
 in magnitude because $t_1^2\simeq 1$.
Conversely, although atmospheric neutrino oscillations can be nearly maximal,
the possibility that $t_1$ is exactly one is excluded.
 The appearance of the large
 factor  $\tan{2\theta_1}$ in Eq. \eB{c}\  requires $s_2\,\cos{\delta}$
 to be  tiny if  $R_\nu$ is to be small.  Thus $s_2$ is unlikely to be
measureable {\it unless\/} $0<|\cos{\delta}|\ll 1$.  If this texture
is correct, and if $s_2$ is found to depart significantly from zero,  CP
violation in the neutrino sector must  be nearly maximal.

This texture is also promising in regard to
the search
for  neutrinoless double beta decay. We obtain 
\eqn\eBB{{\cal M}_{ee}\simeq  -t_1^2\,
\sqrt{\Delta_a/|1-t_1^4|}\,.} 
Because atmospheric neutrino oscillations are observed to be nearly maximal, 
Eq.~\eBB\ tells us that ${\cal M}_{ee}$ is likely to exceed 100~meV.
The rate of neutrinoless double beta  decay may  approach 
its current experimental upper limit.\medskip
\bigskip\medskip

\noindent{Case $B_2$: ~~
$ \pmatrix{X&0&X\cr 0&X&X\cr X&X&0\cr}$}
\medskip
\noindent  This texture, with  ${\cal M}_{\tau\tau}={\cal M}_{e\mu}=0$, is
described by Eqs.~\eB{a,b,c}\ with $t_1$ replaced by $-1/t_1$. Its
phenomenological consequences are nearly the same  as those of Case $B_1$.
\bigskip\medskip

\noindent{Case $B_3$:  ~~
$ \pmatrix{X&0&X\cr 0&0&X\cr X&X&X\cr}$ \qquad  and \qquad Case $B_4$: ~~
$ \pmatrix{X&X&0\cr X&X&X\cr 0&X&0\cr}$}
\medskip
\noindent 
For Case $B_3$, with  ${\cal M}_{\mu\mu}={\cal M}_{e\mu}=0$, we obtain the
relations
\eqna\eBM
$$\eqalignno{
&m_1\simeq -t_1^2\big(1-s_2(e^{-i\delta}\,t_1+e^{i\delta}/t_1)/t_3\big)
\,m_3\,, &\eBM a\cr
&m_2\simeq -t_1^2\big(1+s_2(e^{-i\delta}\,t_1+e^{i\delta}\,/t_1)\,t_3\big)
\,m_3\,.&\eBM b\cr}$$
The phenomenological implications of Eqs.~\eBM{a,b}\ are substantially
the same as  those of Case $B_1$.  The same is true for Case $B_4$
with ${\cal M}_{\tau\tau}={\cal M}_{e\tau}=0$. It
results in  Eqs.~\eBM{a,b}\ with $t_1$ replaced by $-1/t_1$. Thus the  four
 $B_i$ cases are experimentally almost  indistinguishable. 

\vfill\eject

\noindent {Case $C$: ~~
$ \pmatrix{X&X&X\cr X&0&X\cr X&X&0\cr}$}
\medskip
\noindent Our seventh and last  allowed texture has 
 ${\cal M}_{\mu\mu}={\cal M}_{\tau\tau}=0$.
We obtain the relations
\eqna\eC
$$\eqalignno{
&m_1\simeq
-\big(1-e^{i\delta}\,\cot{2\theta_1}/(t_3 s_2)\big)\,m_3\,,& \eC a\cr
&m_2\simeq
-\big(1+e^{i\delta}\,\cot{2\theta_1}\,t_3/ s_2\big)\,m_3\,.& \eC b\cr
}$$
We can obtain a small value of $R_\nu$ 
 if and only if 
$\big\vert \vert m_1\vert -\vert m_2\vert\big\vert \ll m_3$, in
which event we find 
\eqn\eCC{s_2\,\cos{\delta}\approx \cot{2\theta_1}\,\cot{2\theta_3}\,.}
This approximate equality shows that
 $s_2$ can be large enough to be measured if atmospheric neutrino
oscillations are not too nearly  maximal. 
The observed value of
$R_\nu$  results from a small departure from the equality
of  Eq.~\eCC\ by an amount of order
$s_2^2$.

Furthermore, for  Case $C$ we obtain
$$\vert m_{1,2}\vert^2 \simeq  (1+\cos^2{\delta}\,\tan^2{2\theta_3})\,m_3^2
\qquad{\rm and}\qquad |{\cal M}_{ee}|\simeq m_3\,.$$
From these results,  we find
\eqn\eCCC{ |{\cal M}_{ee}| \simeq  \big\vert \sqrt{\Delta_a}\,
\cot{2\theta_3}/\cos{\delta}\big\vert\,.}
Thus  the effective parameter governing neutrinoless double beta decay
must be  at least of order 30~meV and could be considerably larger. 
\medskip

An interesting version of Case~C arises if $\theta_1 \equiv \pi/4$ and $\theta_2 \equiv 0$.
 In place of Eqs.~\eC{a,b}\ we get the single constraint
\eqn\con{s_3^2 m_1+c_3^2 m_2+m_3=0\,.}
With $|m_1|\simeq |m_2|$ to ensure that $R_{\nu}$
is small, Eq.~\con\ yields the relation
$${ {|{\cal M}_{ee}|^2\over \Delta_a }\simeq  {{1-\sin^2 2\theta_3\, \sin^2 {\chi \over 2}} \over 
\sin^2 2\theta_3\, \sin^2 {\chi \over 2}}\,,}$$
where $\chi$ is the relative phase between $m_1$ and $m_2$ and $0<\chi\leq \pi$. Thus
$|{\cal M}_{ee}|$ is bounded below by $\sqrt{\Delta_a}\,\cot{2\theta_3}$, as it is in Eq.~\eCCC.

No other two-zero texture of the neutrino mass matrix is compatible with our
empirical hypotheses. It is easily verified that no two of our allowed
two-zero textures can  be simultaneously
satisfied while remaining  consistent with our
empirical hypotheses.
It follows  that there is no tolerable 
three-zero texture.  For example, a neutrino mass matrix with vanishing
diagonal entries (the Zee Ansatz) cannot yield a large enough value of
$R_\nu$ unless solar neutrino oscillations are very nearly 
maximal~\ref\rfram{P.H. Frampton, M.C. Oh and T. Yoshikawa,
{\tt hep-ph/0110300}.}, a
situation that appears to be strongly disfavored by
experiment.
\vfill\eject

\newsec{Concluding Comment}

The seven  allowed two-zero neutrino textures  fall into three classes:
$A$ (with two members), $B$ (with four members), and $C$.
 The textures within  each class are  difficult 
or impossible to distinguish
experimentally, but each of the three classes has
radically different implications. 
For class $A$, the subdominant angle
$\theta_2$ is expected to be relatively large, but  no-neutrino
$\beta\beta$ decay  is forbidden. For class $B$,  the latter process
should be measureable by the next generation of double beta decay
experiments, whilst $s_2$ may or may not be large enough to be detected.
However, if $s_2$ is comparable to its experimental upper limit, CP violation
must be nearly maximal and should be readily detectable by 
proposed experiments.  For
class  $C$,  no-neutrino
$\beta\beta$ decay is likely to be observable  and
$\theta_2$ ought to  be large enough to 
measure (unless $\theta_1 \equiv \pi/4$) and to permit a search for CP violation. 
It is surprising that such a great variety of textures of the neutrino mass
matrix can fit what is presently known about neutrino masses and
oscillations. 
Future data should reveal
which, if any, of these textures should serve as a guide to the model
builder.

\medskip

We are aware that specific textures  of the neutrino mass
such as we discuss cannot be preserved to all orders in the unspecified
interactions which generate neutrino masses. However, straightforward
elaborations of the standard model can generate zeroes  of $\cal M$
in tree approximation in such a manner that these entries remain tiny
when radiatively corrected. To illustrate this fact we close
with  just
one  example. We sketch  a
 model in which softly-broken lepton number violation  yields  
$\cal M$ with the texture of Case $C$. 
Similar constructions, perhaps more contrived, 
can generate mass matrices with any of  the other textures we have discussed:
      \medskip

In the  basis of charged lepton mass eigenstates,
we introduce two singly-charged SU(2) singlet spinless mesons
 coupled antisymmetrically to pairs of left-handed leptons: $\phi_{e\mu}^-$
to the $e$ and $\mu$ doublets, and $\phi_{\tau\mu}^-$ to the $\tau$ and $\mu$
doublets. An additional
doubly-charged  spinless
singlet, $\chi_{\mu\tau}^{--}$, couples to the
right-handed $\mu^-\,\tau^-$ pair.  These Yukawa couplings conserve lepton
number and flavor symmetry. Soft  breaking of both of these symmetries
 results from 
dim-3 couplings among the various $\phi,\,\phi,\, \chi$ combinations.
Neutrino masses arise from  finite 2-loop diagrams involving
this trilinear coupling and two mass
insertions. The resulting  mass matrix has  
${\cal M}_{\mu\mu}={\cal M}_{\tau\tau}=0$, as desired.

\vfill\eject
\bigskip\bigskip\bigskip
\centerline{\bf Acknowledgements}\bigskip\bigskip

The work of PHF was supported in part by the Department of Energy under grant
number DE-FG02-97ER-41036; that of SLG by the National Science Foundation
under grant number NSF-PHY-0099529; and that of DM by the Department of
Energy under grant number DE-FG02-91ER-40676.

\listrefs
\bye